\documentclass[aps,twocolumn,pra,showpacs,tightenlines]{revtex4}
\usepackage{amssymb}
\usepackage{amsmath}
\usepackage{graphicx}
\usepackage{epsfig}
\usepackage{txfonts}
\usepackage{subfigure}
\usepackage{amsfonts}

\begin{document}
\title{Single particle machine for quantum thermalization}
\author{Jie-Qiao Liao}
\affiliation{Institute of Theoretical Physics, Chinese Academy of
Sciences, Beijing 100190, China}
\author{H. Dong}
\affiliation{Institute of Theoretical Physics, Chinese Academy of
Sciences, Beijing 100190, China}
\author{C. P. Sun}
\affiliation{Institute of Theoretical Physics, Chinese Academy of
Sciences, Beijing 100190, China}

\date{\today}

\begin{abstract}
The long time accumulation of the \textit{random} actions of a
single particle ``reservoir" on its coupled system can transfer some
temperature information of its initial state to the coupled system.
This dynamic process can be referred to as a quantum thermalization
in the sense that the coupled system can reach a stable thermal
equilibrium with a temperature equal to that of the reservoir. We
illustrate this idea based on the usual micromaser model, in which a
series of initially prepared two-level atoms randomly pass through
an electromagnetic cavity. It is found that, when the randomly
injected atoms are initially prepared in a thermal equilibrium state
with a given temperature, the cavity field will reach a thermal
equilibrium state with the same temperature as that of the injected
atoms. As in two limit cases, the cavity field can be cooled and
``coherently heated" as a maser process, respectively, when the
injected atoms are initially prepared in ground and excited states.
Especially, when the atoms in equilibrium are driven to possess some
coherence, the cavity field may reach a higher temperature in
comparison with the injected atoms. We also point out a possible
experimental test for our theoretical prediction based on a
superconducting circuit QED system.
\end{abstract}

\maketitle
\narrowtext

\section{\label{Sec:1}Introduction}
A small system in contact with a large reservoir (or so-called heat
bath) in thermal equilibrium of temperature $T$ will dynamically
approach to an equilibrium state with the same temperature
$T$~\cite{Breuer}. This irreversible process from a nonequilibrium
state into a stable one is conventionally referred to as quantum
thermalization. Most recently, another kind of thermalization,
called canonical thermalization (e.g.,
Refs.~\cite{Popescu,Goldstein,Gemmer,Dong}), investigated in the
meaning of typicality that almost all pure states in the universe
(the system plus its bath) are entangled, and thus the system can
reach an approximately canonical thermal state by averaging over the
bath. Here, the temperature appears as an ``emergent" concept.

In conventional thermalization, the heat bath consists of a very
large number of degrees of freedom (for example, a set of harmonic
oscillators for the bosonic heat bath), and the coupling strengths
of the thermalized system with the degrees of freedom of its bath
are \textit{randomly} distributed. According to the viewpoint in
statistical mechanics that an average over an ensemble is equivalent
to the time average in some sense~\cite{Louisell}, a natural
question is if a series of \textit{random} actions of a
single-particle ``reservoir" injected randomly in a time domain can
transfer some temperature information of its initial state to the
coupled system at a steady state as a thermalization process? To
answer this question, in this paper we study the steady state of a
quantum system which is controlled to have a randomly ``multipulse"
type interaction with a single-particle system initially prepared in
thermal equilibrium with a temperature. If the steady state of the
quantum state is a thermal one with the same temperature as that of
the single-particle system, we think that this quantum system has
been thermalized by the single-particle system through a randomly
``multipulse" type interaction.

Since the randomly ``multipulse" type interaction can be realized by
random injections, in this paper we will illustrate our idea based
on the usual micromaser model (e.g.,
Refs.~\cite{Scully-maser,Meystre,Orszag,Scully,Meschede,Casagrande,Cresser,Cresser1996,Bergou,Guerra,Benkert1993,Bergou1994,Brune,Rempe}),
in which a series of initially prepared atoms pass through an
electromagnetic cavity. Here, the single-mode cavity field is the
system to be thermalized and the randomly injected atoms play the
role of the single-particle reservoir~\cite{Davidovich2007}. Under
some conditions we will clarify, if the injected atoms is initially
prepared in thermal equilibrium, that the conventional
thermalization enables the cavity field to transit from any initial
state to a thermal state with the same temperature as that of the
atoms. We also find that the temperature of the cavity field in
thermal equilibrium depends on the initial state of the injected
atoms. As in two limit cases, such quantum thermalization can
describe the cooling~\cite{Zhang} and masering
processes~\cite{Scully-maser,Meystre,Orszag,Scully,Meschede,Casagrande,Cresser,Cresser1996,Bergou,Guerra,Benkert1993,Bergou1994,Brune,Rempe,You2},
which respectively correspond to the cases where the injected atoms
are initially prepared in ground and excited states.

It is worth noting that when the atoms initially possess some
quantum coherence~\cite{McGowan,Krause}, the ``thermalized state" of
the cavity field will carry the information of this coherence.
Actually, quantum coherence has been proved to be a kind of resource
to enhance quantum information processing. Most recently, some
studies have shown that physical processes with quantum coherence
usually possess some novel effect for energy
transfer~\cite{Fleming2007,Flemingnature}. For example, quantum heat
engines using quantum matter (and even with the assistance of
Maxwell's demon) as a working substance can improve work extraction
as well as the working efficiency in the thermodynamics
cycle~\cite{Scully-science,QPRE2006,QPRL2006}. In the present study,
it is expected that, when the injected two-level atoms possess some
coherence in some situations, the cavity field will reach a steady
state with higher temperature than that for the incoherent case.

Though we calculate the steady-state photon number in the cavity of
the micromaser, we still emphasize that the motivation of this paper
is not to simply study the statistical properties of the cavity
field, but to study the quantum thermalization of a quantum system
randomly coupled to a series of single-particle reservoirs in a time
domain. Therefore our present work is different from other previous
papers on quantum statistical properties of a micromaser (e.g.,
Refs.\cite{Scully-maser,Meystre,Orszag,Scully,Meschede,Casagrande,Cresser,Cresser1996,Bergou,Guerra,Benkert1993,Bergou1994}).
Here we employ the micromaser model only for convenience. The
micromaser involves the process of a quantum system (the single-mode
cavity field) randomly coupled with a series of single-particle
reservoirs (these injected atoms). In other words, the micromaser
model is the platform to show our idea of thermalization. More
importantly, we focus on the \textit{temperature} of the cavity
field at a steady state. If the temperature of the cavity field at a
steady state is equal to that of the injected atoms, we consider
that the cavity field of the micromaser has been thermalized by
these injected atoms. For the case of random injections, the steady
state of the cavity field can be naturally identified as a thermal
state with the same temperature as that of the injected atoms.

In addition, from the viewpoint of experimental implementation, this
paper also provides a possibility for the examination of
thermodynamics with a cavity QED system. As we know, the cavity QED
system has became a mature candidate for the implementation of
experiments in quantum physics and quantum information
processing~\cite{Haroche2001}. Therefore the present work can also
be considered an example for the experimental examination of
thermodynamics with a cavity QED system.

The paper is organized as follows. In Sec.~\ref{Sec:2}, we present
our thermalization model of a single-mode cavity field interacting
with a series of atoms injected randomly. A quantum master equation
is derived to describe the dynamics of the single-mode cavity field.
In Sec.~\ref{Sec:3}, we show that the present quantum thermalization
model can give a unified description of cooling, masering, and
thermalization processes. In Sec.~\ref{Sec:4}, we study the quantum
thermalization when the initial state of the injected two-level
systems possesses some quantum coherence. In Sec.~\ref{Sec:5}, we
propose an experimental implementation of our quantum thermalized
model with superconducting circuit-QED. We also show that the
dynamics of the cavity field in the micromaser is equivalent to the
dynamics of the transmission line resonator in the circuit QED.
Finally, we conclude this paper with some discussions in
Sec.~\ref{Sec:6}.

\section{\label{Sec:2}Cavity QED model for thermalization with single-particle reservoir}
\begin{figure}[tbp]
\includegraphics[bb=8 32 452 266, width=3.2 in]{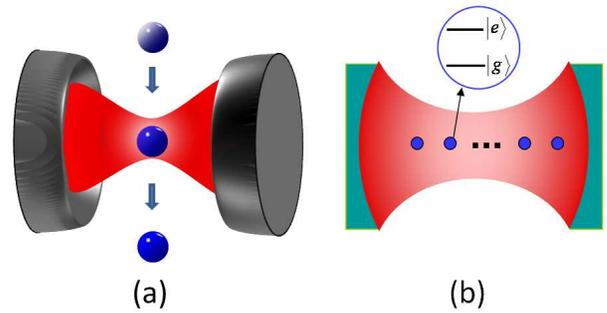}
\caption{(Color online). (a) Schematic diagram of our single
particle thermalization model that a series of prepared TLSs
randomly pass through a single-mode cavity one by one, in
equilibrium it is equivalent to the conventional reservoir model,
(b) where many identical atoms with spatially random distribution
thermalize the single-mode cavity field.} \label{fig1}
\end{figure}

The cavity QED model (as illustrated in Fig.~\ref{fig1}(a)) for
thermalization contains a single-mode cavity field of frequency
$\omega $ and a series of injected two-level systems (TLSs) with
excited state $|e\rangle $, ground state $|g\rangle $, and energy
separation $\omega _{0}$.
These TLSs pass through the single-mode cavity one by one randomly.
Here, the single-mode cavity field is considered as the system to be
thermalized, while the TLSs are considered as the single particle
reservoir. The injections of the TLSs into the cavity are
\textit{random} and there is a limit of one TLS in the cavity each
time. According to the viewpoint in statistical mechanics, the
average over an ensemble is equivalent to the time average in some
sense. It is expected that the single-model cavity field will
approach a steady state equilibrium with a temperature as that of
the injected atoms, since this system is equivalent to the
conventional thermalization model, as shown in Fig.~\ref{fig1}(b),
where many identical atoms (reservoir) with spatially random
distribution thermalize the single-mode cavity field.

A single TLS interacting with the
single-mode cavity field is described by the Jaynes-Cummings (JC)
Hamiltonian
\begin{equation}
\hat{H}=\frac{\omega_{0}}{2}\hat{\sigma}_{z}+\omega
\hat{a}^{\dag}\hat{a}+g(\hat{a}\hat{\sigma}_{+}
+\hat{\sigma}_{-}\hat{a}^{\dag }),\label{JCHamiltonian}
\end{equation}
where $\hat{a}$ and $\hat{a}^{\dag }$ are, respectively, the
annihilation and creation operators of the single-mode cavity field,
they satisfy the usual bosonic commutation relation
$[\hat{a},\hat{a}^{\dag }]=1$. Hereafter we set $\hbar=1$. The
operators of the TLS are defined as
\begin{eqnarray}
\hat{\sigma}_{+}&=&\hat{\sigma}_{-}^{\dag }=|e\rangle \langle g|,\hspace{0.5 cm}
\hat{\sigma}_{z}=|e\rangle \langle e|-|g\rangle \langle g|.
\end{eqnarray}
The parameter $g$ is the coupling strength of the cavity field with a
TLS.

In the rotating picture with respect to
\begin{equation}
\hat{H}_{0}=\frac{\omega}{2}\hat{\sigma}_{z}+\omega
\hat{a}^{\dag}\hat{a},
\end{equation}
the Hamiltonian becomes
\begin{equation}
\hat{V}_{I}=\frac{\delta}{2}
\hat{\sigma}_{z}+g(\hat{a}\hat{\sigma}_{+}
+\hat{\sigma}_{-}\hat{a}^{\dag }),\label{Hininter}
\end{equation}
where
\begin{equation}
\delta\equiv\omega_{0}-\omega
\end{equation}
is the detuning of the cavity
frequency $\omega$ with the energy separation $\omega_{0}$ of the
TLS. In the resonant case, namely $\delta=0$, the unitary evolution
operator governed by the Hamiltonian~(\ref{Hininter}) of the cavity QED reads~\cite{Orszag}
\begin{eqnarray}
\hat{U}(\tau )&\equiv&\exp(-i\hat{V}_{I}\tau)\nonumber\\
&=&\left(
\begin{array}{cc}
\cos\left(g\tau \sqrt{\hat{a}\hat{a}^{\dag }}\right)&-i\frac{\sin\left(g\tau \sqrt{\hat{a}\hat{a}^{\dag }}\right)}{\sqrt{\hat{a}\hat{a}^{\dag }}}\hat{a} \\
-i\hat{a}^{\dag }\frac{\sin\left(g\tau \sqrt{\hat{a}\hat{a}^{\dag
}}\right)}{\sqrt{\hat{a}\hat{a}^{\dag }}} & \cos \left(g\tau
\sqrt{\hat{a}^{\dag }\hat{a}}\right)
\end{array}
\right),
\end{eqnarray}
which is written in the Hilbert subspace of the TLS with the basis
states
\begin{eqnarray}
|e\rangle=\left(
                  \begin{array}{c}
                    1 \\
                    0 \\
                  \end{array}
                \right),\hspace{0.5 cm}
|g\rangle=\left(
                  \begin{array}{c}
                    0 \\
                    1 \\
                  \end{array}
                \right).
\end{eqnarray}

To thermalize the cavity field, a series of TLSs are randomly
injected into the cavity for a fixed time interval $\tau$. All of
the TLSs are initially prepared in the density matrix
\begin{equation}
\hat{\rho}_{\text{TLS}}=p_{e}|e\rangle \langle e|+p_{g}|g\rangle
\langle g|+\lambda |e\rangle \langle g|+\lambda^{\ast }|g\rangle
\langle e|,\label{atominitialstate}
\end{equation}
where $\lambda$ is the parameter describing the coherence of the
TLSs. The state preparation of the TLSs can be realized by using a
pumping field to excite the TLSs. We assume that the $j$th TLS is
injected into the cavity at time $t_{j}$. After an interaction of
time $\tau$, the state of the cavity field becomes
\begin{eqnarray}
\hat{\rho}(t_{j}+\tau)&=&\texttt{Tr}_{\textrm{TLS}}[\hat{U}(\tau)\hat{\rho}(t_{j})\otimes\hat{\rho}
_{\textrm{TLS}}U^{\dagger }\left( \tau
\right)]\nonumber\\
&\equiv&\mathcal{M}(\tau)\hat{\rho}(t_{j}),\label{superoperatorintro}
\end{eqnarray}
where $\texttt{Tr}_{\textrm{TLS}}$ means tracing over the degree of
freedom of the TLS. The superoperator $\mathcal{M}\left( \tau
\right)$ introduced in Eq.~(\ref{superoperatorintro}) can be
expressed as follows:
\begin{widetext}
\begin{eqnarray}
\mathcal{M}(\tau)\hat{\rho}(t_{j})
&=&p_{e}\cos \left(g\tau \sqrt{\hat{a}\hat{a}^{\dag }}\right)
\hat{\rho}(t_{j})\cos \left(g\tau \sqrt{\hat{a}\hat{a}^{\dag
}}\right)+p_{e}\hat{a}^{\dag }\frac{\sin\left(g\tau
\sqrt{\hat{a}\hat{a}^{\dag }}\right)}{\sqrt{\hat{a}\hat{a}^{\dag
}}}\hat{\rho}(t_{j})\frac{\sin\left(g\tau \sqrt{\hat{a}\hat{a}^{\dag
}}\right)}{\sqrt{\hat{a}\hat{a}^{\dag }}}
\hat{a}\nonumber\\
&&+p_{g}\frac{\sin\left(g\tau \sqrt{\hat{a}\hat{a}^{\dag
}}\right)}{\sqrt{\hat{a}\hat{a}^{\dag
}}}\hat{a}\hat{\rho}(t_{j})\hat{a}^{\dagger }\frac{\sin\left(g\tau
\sqrt{\hat{a}\hat{a}^{\dag }}\right)}{\sqrt{\hat{a}\hat{a}^{\dag
}}}+p_{g}\cos \left(g\tau \sqrt{\hat{a}^{\dag
}\hat{a}}\right)\hat{\rho}(t_{j})\cos \left(g\tau
\sqrt{\hat{a}^{\dag }\hat{a}}\right)\nonumber\\
&&+i\lambda \cos \left(g\tau \sqrt{\hat{a}\hat{a}^{\dag
}}\right)\hat{\rho}(t_{j})\hat{a}^{\dagger }\frac{\sin\left(g\tau
\sqrt{\hat{a}\hat{a}^{\dag }}\right)}{\sqrt{\hat{a}\hat{a}^{\dag
}}}-i\lambda \hat{a}^{\dag }\frac{\sin\left(g\tau
\sqrt{\hat{a}\hat{a}^{\dag }}\right)}{\sqrt{\hat{a}\hat{a}^{\dag
}}}\hat{\rho}(t_{j})\cos \left(g\tau
\sqrt{\hat{a}^{\dag}\hat{a}}\right)\nonumber\\
&&+i\lambda ^{\ast }\cos \left(g\tau \sqrt{\hat{a}^{\dag
}\hat{a}}\right)\hat{\rho}(t_{j})\frac{\sin\left(g\tau
\sqrt{\hat{a}\hat{a}^{\dag }}\right)}{\sqrt{\hat{a}\hat{a}^{\dag
}}}\hat{a}-i\lambda ^{\ast }\frac{\sin\left(g\tau
\sqrt{\hat{a}\hat{a}^{\dag }}\right)}{\sqrt{\hat{a}\hat{a}^{\dag
}}}\hat{a}\hat{\rho}(t_{j})\cos\left(g\tau
\sqrt{\hat{a}\hat{a}^{\dag }}\right),\label{msuperoperator}
\end{eqnarray}
\end{widetext}

In fact, in addition to the action of the injected TLSs, the cavity
inevitably couples with an external environment through the cavity wall.
Within the quantum noise theory, we model the external
environment of the cavity as a heat bath. When the
coupling of the cavity field with the heat bath is weak, the decay of the cavity field can be described by
~\cite{Scully}
\begin{eqnarray}
\mathcal{L}\hat{\rho}&=&\frac{1}{2}\kappa
(\bar{n}_{th}+1)(2\hat{a}\hat{\rho}\hat{a}^{\dag }-\hat{a}^{\dag
}\hat{a}\hat{\rho}-\hat{\rho}\hat{a}
^{\dag }\hat{a})  \notag\\
&&+\frac{1}{2}\kappa \bar{n}_{th}(2\hat{a}^{\dag
}\hat{\rho}\hat{a}-\hat{a} \hat{a}^{\dag
}\hat{\rho}-\hat{\rho}\hat{a}\hat{a}^{\dag
}),\label{dissipationoperator}
\end{eqnarray}
where $\kappa$ is the decay rate of the cavity. The
thermal average photon number is
\begin{eqnarray}
\bar{n}_{th}=\frac{1}{e^{\beta_{b}\omega}-1}
\end{eqnarray}
with $\beta_{b}=1/(k_{B}T)$ being the inverse temperature of the
heat bath. Hereafter we denote $\hat{\rho}(t)$ as $\hat{\rho}$ to be
concise.

Since the TLSs are injected at \textit{random}, we can introduce a rate $r$ of a Poisson process to
depict the arrival of the TLSs. In a time interval of  $(t,t+\delta t)$, the probability
of a TLS arrival is $r\delta t$. Hence the density matrix of the cavity field at time $t+\delta t$ can be written as~\cite{comments}
\begin{eqnarray}
\hat{\rho}(t+\delta t)=(1-r\delta t)\left[\hat{\rho}(t)+\mathcal{L}\hat{\rho}(t)\delta t\right]+r\delta t\mathcal{M}(\tau)\hat{\rho}(t).
\label{diffequation}
\end{eqnarray}
Here the first term on the right-hand side of
Eq.~(\ref{diffequation}) describes the density matrix of the cavity
field at time $t+\delta t$ when a TLS does not pass through the
cavity, with the probability $1-r\delta t$. In this case, the cavity
field evolves under the action of $\mathcal{L}$. Additionally, the
last term on the right-hand side of Eq.~(\ref{diffequation})
describes the density matrix of the cavity field at time $t+\delta
t$ for the case of a TLS passing through the cavity, with the
probability $r\delta t$. Notice that here we approximately neglect
the action of the heat bath on the cavity field during the process
of the TLS passing through the cavity, since the time spent by each
TLS in the cavity is assumed to be much shorter than the mean time
between two injections of the TLSs.

Taking the limit of $\delta t\rightarrow0$, we can obtain the
following quantum master
equation~\cite{Orszag,Scully,Meschede,Casagrande,Cresser,Cresser1996,Bergou,Guerra}
\begin{equation}
\dot{\hat{\rho}}=r(\mathcal{M}(\tau)-1)\hat{\rho}
+\mathcal{L}\hat{\rho}\label{masterequation}
\end{equation}
to describe the evolution of the single-mode cavity field.

\section{\label{Sec:3}Unification of cooling, masering and thermalization}
The evolution of the quantum state of the cavity field is governed
by the master equation (\ref{masterequation}), which depends on the
initial state of the injected TLSs. Firstly, we consider the
case where no coherence exists in the initial state of
the TLSs, i.e., $\lambda =0$ in Eq.~(\ref{atominitialstate}). In the
Fock state representation, the evolution equation for the diagonal
elements $P_{n}=\langle n|\hat{\rho}|n\rangle $ of the density
matrix $\hat{\rho}$ in the master equation~(\ref{masterequation})
becomes
\begin{eqnarray}
\dot{P}_{n}&=&rp_{e}\left[\cos^{2}\left(g\tau\sqrt{n+1}\right)P_{n}
+\sin^{2}\left(g\tau\sqrt{n}\right)P_{n-1}\right]\nonumber\\
&&+rp_{g}\left[\cos^{2}\left(g\tau\sqrt{n}\right)P_{n}
+\sin^{2}\left(g\tau\sqrt{n+1}\right)P_{n+1}\right]\nonumber\\
&&-rP_{n}+\frac{\kappa\bar{n}_{th}}{2}\left[2nP_{n-1}-2(n+1)P_{n}\right]\nonumber\\
&&+\frac{\kappa(\bar{n}_{th}+1)}{2}\left[2(n+1)P_{n+1}-2nP_{n}\right].\label{elementsequation}
\end{eqnarray}
Using the relation $p_{e}+p_{g}=1$ and after some simple collection,
the above equation (\ref{elementsequation}) becomes
\begin{eqnarray}
\dot{P}_{n}&=&-r\sin^{2}\left(g\tau\sqrt{n+1}\right)(p_{e}P_{n}-p_{g}P_{n+1})\nonumber\\
&&-\kappa(n+1)\left[\bar{n}_{th}P_{n}-(\bar{n}_{th}+1)P_{n+1}\right]\nonumber\\
&&+r\sin^{2}\left(g\tau\sqrt{n}\right)(p_{e}P_{n-1}-p_{g}P_{n})\nonumber\\
&&+\kappa n\left[\bar{n}_{th}P_{n-1}-(\bar{n}_{th}+1)P_{n}\right].
\end{eqnarray}
The steady state solution $\dot{P}_{n}=0$ leads to the detailed
balance condition and the relation
\begin{eqnarray}
&&r\sin^{2}\left(g\tau\sqrt{n}\right)(p_{e}P_{n-1}-p_{g}P_{n})\nonumber\\
&&+\kappa n\left[\bar{n}_{th}P_{n-1}-(\bar{n}_{th}+1)P_{n}\right]=0.
\end{eqnarray}
Then the ratio $R_{n}=P_{n}/P_{n-1}$ between two neighboring photon
number populations is obtained as
\begin{equation}
R_{n}=\frac{rp_{e}\sin^{2}\left(g\tau \sqrt{n}\right)+\kappa
\bar{n}_{th}n}{ rp_{g}\sin^{2}\left(g\tau \sqrt{n}\right)+\kappa
(\bar{n}_{th}+1)n}.\label{pratio}
\end{equation}
We can understand such thermalization to the steady state with the
definite population ratio (\ref{pratio}) as a temperature
information transfer process from the TLSs to the cavity field,
namely, the curve of $-(\ln R_{n})/\omega$ can explicitly reflect
the information of the temperature of the TLSs.

Such a temperature information transfer process can result in
various coherent manipulations for quantum state engineering. An
example is the cooling of the cavity field as a generalized
thermalization for all injected TLSs initially prepared in the
ground state, i.e., $p_{e}=0$ and $p_{g}=1$. In this case, the TLSs
on the ground state will take away the energy of the cavity field
and then cool it to reach a lower temperature defined by the
decreased photon population
\begin{equation}
P_{n}=P_{0}\prod_{l=1}^{n}\frac{\bar{n}_{th}l}{(\bar{n}_{th}+1)l+\sin^{2}\left(g\tau\sqrt{l}\right)r/\kappa},
\end{equation}
where $P_{0}$ is determined by the normalization condition
$\sum_{n=0}^{\infty}P_{n}=1$. This generalized thermalization
mechanism was even used to cool the nanomechanical resonator by the
pulse-driven charge qubit~\cite{Zhang}. Another example with
$p_{e}=1$ and $p_{g}=0$ shows the maser processes of the cavity
field~\cite{Scully-maser,Meystre,Orszag,Scully,Meschede,Casagrande,Cresser,Cresser1996,Bergou,Guerra,Benkert1993,Bergou1994,Brune,Rempe,You2},
which is represented by the amplified photon population
\begin{equation}
P_{n}=P_{0}\prod_{l=1}^{n}\frac{\sin^{2}\left(g\tau
\sqrt{l}\right)r/\kappa+ \bar{n}_{th}l}{(\bar{n}_{th}+1)l},
\end{equation}
where $P_{0}$ is determined by the normalization condition
$\sum_{n=0}^{\infty}P_{n}=1$.
\begin{figure}[tbp]
\includegraphics[width=8 cm]{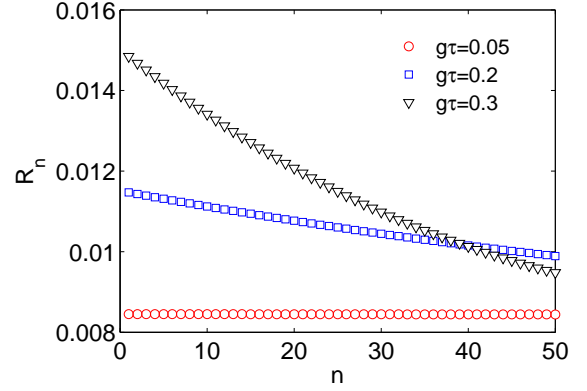}
\caption{(Color online). The ratio $R_{n}$ versus the photon number
$n$ is plotted for $g\tau=0.05$, $0.2$, and $0.3$. The injected TLSs
are prepared in thermal equilibrium of inverse temperature
$\beta=2.898$ ($T=200$ mK). Other parameters are set as $\omega=1$,
$\kappa/\omega=10^{-4}$, $r=2\times10^{-4}$, and
$\bar{n}_{th}=0.008$ ($T_{b}=100$ mK).}\label{ratio}
\end{figure}

In the absence of the cavity field dissipation, i.e., $\kappa =0$,
Eq.~(\ref{pratio}) becomes $R_{n}=p_{e}/p_{g}$, which is
irrespective of both the index $n$ and the average injection rate
$r$. For this case, the temperature information of the TLS is
perfectly transferred to the cavity field. For example, when the TLS
is initially prepared in the thermal equilibrium with temperature
$T$, that is
\begin{eqnarray}
p_{e}(T)=\frac{\exp(-\beta\omega /2)}{2\cosh(\beta
\omega/2)},\hspace{0.5 cm} p_{g}(T)=\frac{\exp(\beta\omega
/2)}{2\cosh(\beta \omega /2)},
\end{eqnarray}
where $T=1/(k_{B}\beta)$, then the population ratio $R_{n}=\exp
(-\beta \omega )$ of the cavity field is independent of the index
$n$, thus a thermal equilibrium has the same temperature $T$ as that
of the TLS.

We give a physical explanation about the steady state of the cavity
field in the absence of the cavity decay. When $\kappa=0$, the
evolution of the cavity is governed by the following quantum master
equation
\begin{equation}
\dot{\hat{\rho}}=r(\mathcal{M}(\tau)-1)\hat{\rho}.\label{masterequationforzerok}
\end{equation}
In the short $\tau$ case, we can
make the short time approximation,
\begin{subequations}
\label{approximation}
\begin{align}
\cos \left(g\tau \sqrt{\hat{a}\hat{a}^{\dag }}\right)&\approx
1-(g\tau
)^{2}\hat{a}\hat{a}^{\dag }/2,\\
\cos \left(g\tau \sqrt{\hat{a}^{\dag }\hat{a}}\right)&\approx
1-(g\tau
)^{2}\hat{a}^{\dag }\hat{a}/2,\\
\sin \left(g\tau \sqrt{\hat{a}\hat{a}^{\dag }}\right)&\approx g\tau
\sqrt{\hat{a}\hat{a}^{\dag }}.
\end{align}
\end{subequations}
Up to the second order of $\tau$, the master
equation~(\ref{masterequationforzerok}) becomes
\begin{eqnarray}
\dot{\hat{\rho}}&\approx&\frac{\alpha p_{g}}{2}(2\hat{a}\hat{\rho}
\hat{a} ^{\dag }-\hat{\rho} \hat{a}^{\dag }\hat{a}-\hat{a}^{\dag
}\hat{a}\hat{\rho}
)\nonumber\\
&&+\frac{\alpha p_{e}}{2}(2\hat{a}^{\dag }\hat{\rho}
\hat{a}-\hat{\rho} \hat{a}\hat{a}^{\dag }-\hat{a}\hat{a}^{\dag
}\hat{\rho}),\label{effmasequforkaeqzero}
\end{eqnarray}
where $\alpha =r(g\tau )^{2}$. Now, the injected TLSs are prepared
in a statistical mixture of the excited and ground states. From the
above equation~(\ref{effmasequforkaeqzero}), we can see that the
TLSs prepared in an \textit{excited} state \textit{excite} the
cavity at an effective rate $\alpha p_{e}$, while the TLSs prepared
in a \textit{ground} state \textit{take away} the energy excitation
of the cavity field at an effective rate $\alpha
p_{g}$~\cite{Laserphysics}. By comparing
Eq.~(\ref{effmasequforkaeqzero}) with
Eq.~(\ref{dissipationoperator}), we can see that the long time
accumulation of the actions of the injected TLSs is equivalent to an
effective heat bath with the inverse temperature
\begin{eqnarray}
\beta_{\textrm{eff}}^{th}=-\frac{1}{\omega}\ln\frac{p_{e}}{p_{g}}.
\end{eqnarray}
Therefore, the cavity field can reach a steady state even in the
absence of the cavity decay through the walls. It is of interest
that the inverse temperature $\beta_{\textrm{eff}}^{th}$ of the
effective heat bath of the cavity can be controlled by changing the
populations $p_{e}$ and $p_{g}$ of the injected TLSs.

In the presence of the cavity field dissipation, i.e.,
$\kappa\neq0$, generally, it is impossible to define a temperature
for the cavity field in the steady state, since in this case the
ratio $R_{n}$ given by Eq.~(\ref{pratio}) depends on $n$. In
Fig.~\ref{ratio}, we plot $R_{n}$ versus photon number $n$ for
different $g\tau$. Clearly, for small $g\tau$, $R_{n}$ shows the
independence of the photon number $n$. Therefore, it is possible to
define an effective temperature for the cavity field when $g\tau$ is
small.

In the short interaction time $\tau$ limit, i.e., $g\tau \sqrt{n}\ll
1$ for all experimental accessible photon numbers $n$, we make an
approximation $\sin^{2}\left(g\tau\sqrt{n}\right)\approx (g\tau
)^{2}n$, which results in an $n-$independent population ratio
\begin{equation}
R=\frac{\alpha p_{e}+\kappa \bar{n}_{th}}{\alpha p_{g}+\kappa
(\bar{n}_{th}+1)}.
\end{equation}
Thus for the TLS injection in thermal equilibrium, we can define an
effective inverse temperature for the cavity field
\begin{equation}
\beta _{\text{eff}}=-\frac{1}{\omega }\ln R,
\end{equation}
which satisfies the relation
\begin{equation}
\min \{\beta _{b},\beta \}<\beta _{\text{eff}}<\max \{\beta _{b},\beta \}.
\end{equation}
It means that the cavity field will approach a thermal equilibrium
with an intermediate inverse temperature $\beta _{\text{eff}}$
between those for the TLSs and the heat bath. Additionally, for the
case of $\beta _{b}=\beta $, the cavity field will approach a
thermal equilibrium of $\beta _{\text{eff}}=\beta $. This result is
reasonable from the viewpoint of quantum noise. A system coupled
with two heat baths with different temperatures will reach an
equilibrium with intermediate temperatures between those of the two
heat baths~\cite{zhaonan}.

\section{\label{Sec:4}Quantum coherence assisted thermalization}

In the above section, we study the generalized thermalization for
the incoherent case, in which the injected TLSs do not possess
quantum coherence. In this section, we study the coherent case,
i.e., $\lambda \neq 0$. In this case, up to the second order of
$\tau$ the master equation (\ref{masterequation}) can be reduced to
\begin{equation}
\dot{\hat{\rho}}\approx
i[\hat{\rho},\hat{H}_{\text{eff}}]+\mathcal{J}\hat{\rho}\label{coherentMEq}
\end{equation}
in the short $\tau$ limit, where the effective Hamiltonian reads
\begin{equation}
\hat{H}_{\text{eff}}=\xi \hat{a}^{\dag }+\xi ^{\ast }\hat{a},
\end{equation}
with $\xi =rg\tau \lambda$. The superoperator $\mathcal{J}$ is
defined as
\begin{eqnarray}
\mathcal{J}\hat{\rho}&=&\frac{1}{2}\gamma _{1}(2\hat{a}^{\dag
}\hat{\rho}\hat{a}-\hat{a}\hat{a}^{\dag
}\hat{\rho}-\hat{\rho}\hat{a}\hat{a}^{\dag })
\notag \\
&&+\frac{1}{2}\gamma _{2}(2\hat{a}\hat{\rho}\hat{a}^{\dag
}-\hat{a}^{\dag }\hat{a}\hat{\rho}-\hat{\rho}\hat{a}^{\dag
}\hat{a}),
\end{eqnarray}
where we introduced two transition rates: the decay rate $\gamma
_{1}$ and the excitation rate $\gamma_{2}$,
\begin{subequations}
\begin{align}
\gamma _{1}&=\alpha p_{e}+\kappa \bar{n}_{th},\\
\gamma_{2}&=\alpha p_{g}+\kappa (\bar{n}_{th}+1).
\end{align}
\end{subequations}
During the derivation of the master equation~(\ref{coherentMEq}), we
have used the approximation given in Eq.~(\ref{approximation}).

The above effective Hamiltonian $\hat{H}_{\text{eff}}$ describes the
role of the quantum coherence of the injected TLSs: the off-diagonal
terms in the initial state offer nonvanishing atomic transition,
which is added as a driving source of the cavity field. The
generalized master equation~(\ref{coherentMEq}) describes a driven
cavity field in contact with an effective bath characterized by two
rates. This effective bath consists of a TLS reservoir and a heat
bath. It is worth pointing out that the properties of the TLS
reservoir can be manipulated through changing the initial
populations $p_{g}$ and $p_{e}$.

From the master equation~(\ref{coherentMEq}), we can obtain the
following equation of motion of the average value of the creation,
annihilation, and photon number operators,
\begin{subequations}
\label{eqofmotion}
\begin{align}
\frac{d}{dt}\langle\hat{a}(t)\rangle&=-\frac{1}{2}\left( \gamma
_{2}-\gamma _{1}\right) \left\langle
\hat{a}\left( t\right) \right\rangle -i\xi,\\
\frac{d}{dt}\langle\hat{a}^{\dagger}(t)\rangle&=-\frac{1}{2}\left(
\gamma _{2}-\gamma _{1}\right)\langle \hat{a}^{\dagger
}\left( t\right)\rangle +i\xi^{\ast },\\
\frac{d}{dt}\langle\hat{n}(t)\rangle&=-\left( \gamma _{2}-\gamma
_{1}\right) \left\langle \hat{n}\left( t\right) \right\rangle
-i\xi\langle \hat{a}^{\dagger }\left( t\right)\rangle +i\xi^{\ast
}\left\langle \hat{a}\left( t\right) \right\rangle+\gamma_{1}.
\end{align}
\end{subequations}
The steady state solutions of the above equation are
\begin{subequations}
\begin{align}
\langle \hat{a}\rangle_{ss} &=\langle
\hat{a}^{\dag}\rangle_{ss}^{\ast}=-\frac{
2i\xi}{\gamma _{2}-\gamma _{1}},\\
\left\langle \hat{n}\right\rangle_{ss} &=\frac{4\left\vert
\xi\right\vert ^{2}}{\left( \gamma _{2}-\gamma _{1}\right)
^{2}}+\frac{\gamma _{1}}{\gamma _{2}-\gamma
_{1}}.\label{averageofnumber}
\end{align}
\end{subequations}
The above results show that the quantum coherence can increase the
steady state average photon number (with the first term in
Eq.~(\ref{averageofnumber})) in the cavity field. When no TLS is
injected, i.e., $r=0$, the average photon number at steady state
$\langle \hat{n}\rangle _{ss}=\bar{n}_{th}$. On the other hand, for
the case of thermal TLSs injection of inverse temperature $\beta $
and a perfect cavity, i.e., $\kappa =0$, the steady state average
photon number $\langle \hat{n}\rangle _{ss}=1/[\exp (\beta \omega
)-1]$, which implies that the cavity approaches an equilibrium with
the same temperature of the TLSs.
\begin{figure}[tbp]
\includegraphics[width=8 cm]{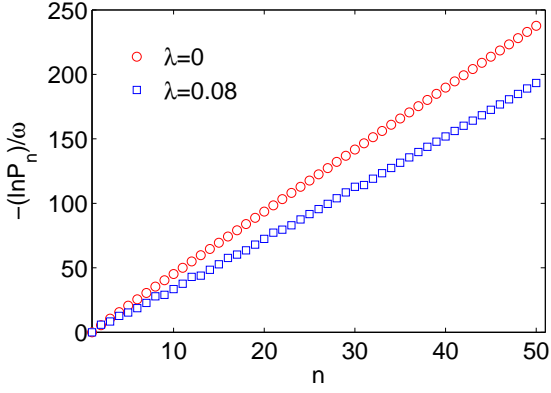}
\caption{(Color online). The $-(\ln P_{n})/\omega$ versus the photon
number $n$ is plotted for both the coherent ($\lambda=0.08$) and
incoherent ($\lambda=0$) cases. Here, $g\tau=0.05$ and other
parameters are set as those in Fig.~\ref{ratio}.}
\label{Nocoherence}
\end{figure}

The above argument based on the short time approximation is only a
heuristic analysis, thus we need to numerically solve the master
equation~(\ref{masterequation}) directly. In Fig.~\ref{Nocoherence},
we plot the $-(\ln P_{n})/\omega$ versus the photon number $n$ for
the coherence and incoherence cases. It can be seen from
Fig.~\ref{Nocoherence} that the coherence in the initial state of
the TLSs can increase the steady state temperature of the cavity.
However, for the coherence case, the dependence of $-(\ln
P_{n})/\omega$ on the photon number $n$ is approximately linear
therefore it is an approximation to define an effective temperature
for the cavity field.

\section{\label{Sec:5}Experimental implementation with circuit QED system}

\begin{figure}[tbp]
\includegraphics[bb=68 583 390 756, width=7.5 cm]{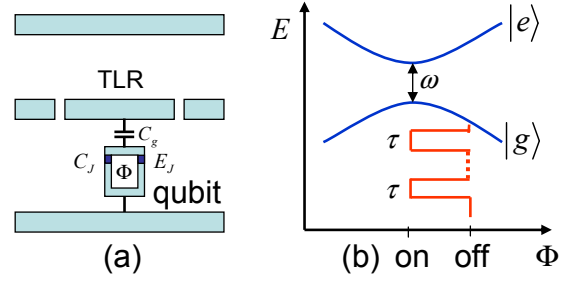}
\caption{(Color online). (a) Schematic diagram of the circuit QED
system of a transmission line resonator (TLR) coupled with a charge
qubit, which is controlled through an external magnetic flux $\Phi$
and a gate voltage $V$. (b) The energy levels of the charge qubit
versus the magnetic flux $\Phi$, the working status of the qubit is
controlled by the magnetic flux ``pulse series". By tuning the
magnetic flux $\Phi$, we can switch on or off the coupling between
the TLR and the charge qubit.} \label{CircuitQED}
\end{figure}

In this section, we present an experimental implementation of our
quantum thermalization based on superconducting circuit QED
system~\cite{cQED1,cQED2}. As shown in Fig.~\ref{CircuitQED}(a), a superconducting transmission line resonator (TLR) couples to a superconducting charge qubit. After the
quantization of the electromagnetic field in the TLR,
the Hamiltonian describing a single-model field in the TLR reads~\cite{cQED1}
\begin{eqnarray}
\hat{H}_{\textrm{TLR}}=\omega \hat{a}^{\dagger}\hat{a},
\end{eqnarray}
where $\omega$ is the resonant frequency of this mode. Here we only
choose the single mode which is (near) resonant with the lowest two
levels of the superconducting Cooper-pair box, i.e., the charge
qubit. The Hamiltonian of the Cooper-pair box (CPB) superconducting
circuit is~\cite{Makhlin2001}
\begin{eqnarray}
\hat{H}_{\textrm{CPB}} &=&4E_{c}\sum_{n\in \mathbb{Z}}\left(n-n_{g}\right)^{2}\left\vert n\right\rangle
\left\langle n\right\vert-E_{J}\cos\left(\frac{\pi\Phi}{\Phi_{0}}\right)\nonumber\\
&&\times\sum_{n\in \mathbb{Z}}\left(\left\vert
n+1\right\rangle \left\langle n\right\vert +\left\vert n\right\rangle
\left\langle n+1\right\vert \right),\label{Hofcpb}
\end{eqnarray}
where $E_{c}=e^{2}/(2C_{\Sigma})$ is the Coulomb energy, with
$C_{\Sigma}=2C_{J}+C_{g}$ being the total capacitance connected with
the superconducting island. $E_{J}$ is the Josephson coupling energy
of a single Josephson junction. An external magnetic flux $\Phi$
through the loop can tune the effective Josephson coupling energy of
the CPB. The symbol $\Phi_{0}$ is the magnetic flux quanta. In
addition, we introduce the gate Cooper-pair number
$n_{g}=C_{g}V/(2e)$ with the gate capacitance $C_{g}$ and the gate
voltage $V$. The state $|n\rangle$ ($n\in \mathbb{Z}$, where
$\mathbb{Z}$ denotes the integer set) stands for $n$ extra Cooper
pairs on the island.

In the present scheme, since the existence of the TLR, the gate voltage contains two parts: a dc part $V_{g}^{dc}$
and a quantum part $V_{q}$ generated by the TLR. Then
\begin{eqnarray}
n_{g}=\frac{C_{g}(V_{g}^{dc}+V_{q})}{2e}=n_{g}^{dc}+
\frac{C_{g}V_{q}}{2e},\label{gatenumber}
\end{eqnarray}
where $n_{g}^{dc}=C_{g}V_{g}^{dc}/(2e)$. The voltage generated by the TLR can be written as
\begin{eqnarray}
V_{q}=\sqrt{\frac{\omega}{Lc}}(\hat{a}+\hat{a}^{\dagger}),\label{voltage}
\end{eqnarray}
where $L$ is the length of the TLR and $c$ is the capacitance per unit length of the TLR.

With the substitution of Eqs.~(\ref{gatenumber}) and~(\ref{voltage})
into the Hamiltonian~(\ref{Hofcpb}) and restriction into the
subspace with basis states $|0\rangle$ and $|1\rangle$, we obtain
\begin{eqnarray}
\hat{H}_{\textrm{CPB}}&=&-2E_{c}\left( 1-2n_{g}^{dc}\right)\hat{\tau}_{z}-E_{J}\cos\left(\frac{\pi\Phi}{\Phi_{0}}\right)\hat{\tau}_{x}\nonumber\\
&&-e\frac{C_{g}}{C_{\Sigma }}
\sqrt{\frac{\omega}{Lc}}(\hat{a}+\hat{a}^{\dagger})(
1-2n_{g}^{dc}-\hat{\tau} _{z}),
\end{eqnarray}
where we have introduced the Pauli operators
\begin{eqnarray}
\hat{\tau}_{x}=\left\vert 1\right\rangle \left\langle 0\right\vert
+\left\vert 0\right\rangle \left\langle 1\right\vert,\hspace{0.5 cm}
\hat{\tau}_{z}=\left\vert 0\right\rangle \left\langle 0\right\vert
-\left\vert 1\right\rangle \left\langle 1\right\vert,
\end{eqnarray}
and discarded some constant terms.
When the charge qubit is working at the optimal point
$n^{dc}_{g}=1/2$, the total Hamiltonian of the system becomes
\begin{eqnarray}
\hat{H}&=&\omega \hat{a}^{\dagger}\hat{a}-E_{J}\cos\left(\frac{\pi\Phi}{\Phi_{0}}\right)\hat{\tau}_{x}
+e\frac{C_{g}}{C_{\Sigma}}
\sqrt{\frac{\omega}{Lc}}(\hat{a}+\hat{a}^{\dagger})\hat{\tau} _{z}.\nonumber\\
\end{eqnarray}
By making a rotation
\begin{eqnarray}
\hat{\tau}_{x}\rightarrow-\hat{\sigma}_{z},\hspace{0.5 cm}
\hat{\tau}_{z}\rightarrow\hat{\sigma}_{x},
\end{eqnarray}
the Hamiltonian becomes~\cite{cQED1}
\begin{eqnarray}
\hat{H}&=&\omega \hat{a}^{\dagger}\hat{a}+\frac{\omega_{0}}{2}\hat{\sigma}_{z}+g(\hat{a}+\hat{a}^{\dagger})\hat{\sigma}_{x}.\label{cirQEDHamltonian}
\end{eqnarray}
where the energy separation of the charge qubit is
\begin{eqnarray}
\omega_{0}=2E_{J}\cos\left(\frac{\pi\Phi}{\Phi_{0}}\right),\label{energyseparation}
\end{eqnarray}
and the coupling strength is
\begin{eqnarray}
g=e\frac{C_{g}}{C_{\Sigma }}
\sqrt{\frac{\omega}{Lc}}.\label{couplingstrength}
\end{eqnarray}
The Hamiltonian~(\ref{cirQEDHamltonian}) reduces to the usual JC Hamiltonian given in Eq.~(\ref{JCHamiltonian})
by making the rotation wave approximation.

From Eqs.~(\ref{energyseparation}) and~(\ref{couplingstrength}), we
can see that the energy separation $\omega_{0}$ of the charge qubit
is tunable by controlling the biasing magnetic flux $\Phi$, and the
coupling strength $g$ is fixed once the superconducting circuit is
fabricated, therefore the effective method to switch on and off the
coupling of the resonator with the charge qubit is to tune $\Phi$
such that the qubit couples with the resonator in resonant and very
largely detuned, respectively. We schematically plot the qubit
energy level versus the magnetic flux $\Phi$ in
Fig.~\ref{CircuitQED}(b). The working points ``on" and ``off"
correspond respectively to the resonant coupling and decoupling of
the qubit with the cavity. The magnetic flux ``pulse serial"
controls the interaction of the qubit with the resonator. Note that
similar methods have been proposed to generate photon Fock
states~\cite{Liu} and have recently been realized based on a circuit
QED system consisting of a transmission line resonator coupled with
a phase qubit~\cite{Martinis,Martinis2}.

However, it should be emphasized that, strictly speaking, the
present circuit QED system is different from the micromaser system
since a micromaser has many independent atoms, while the TLR has
strictly one TLS. Therefore we need to know the conditions under
which the dynamics of the single-mode field in the TLR is equivalent
to that of the cavity field in the micromaser. Without loss of
generality, in the following we study the dynamics of the circuit
QED during a single cycle. We assume that the coupling of the charge
qubit with the TLR is switched on at time $t_{i}$, and after an
interaction of time $\tau$, this coupling is switched off. We denote
the density matrix of the circuit QED at time $t_{i}+\tau$ as
\begin{eqnarray}
\hat{\rho}_{\textrm{cir-QED}}=\sum_{m,n=0}^{\infty}\sum_{r,s=\{e,g\}}\rho_{mnrs}|m\rangle_{\textrm{TLR}}\langle n|_{\textrm{TLR}}\otimes|r\rangle_{q}\langle s|_{q},\label{denmatrix}
\end{eqnarray}
where states $|m(n)\rangle_{\textrm{TLR}}$ and $|e(g)\rangle_{q}$
denote the states of the TLR and the charge qubit, respectively.
Generally, this density matrix given in Eq.~(\ref{denmatrix}) is an
entangled state due to the coupling between the charge qubit and the
TLR. During the time interval from $t_{i}+\tau$ to $t_{i+1}$, the
time of the $(i+1)$th turning on the coupling, the density
matrix~(\ref{denmatrix}) evolves under the local actions of the
environments of the charge qubit and the TLR.

In a micromaser, correspondingly, we only focus on the quantum state
of the cavity by tracing over the atom. Therefore, if the total
density matrix of the cavity and the $i$th injected atom is
\begin{eqnarray}
\hat{\rho}_{\textrm{cav-QED}}=\sum_{m,n=0}^{\infty}\sum_{r,s=\{e,g\}}\rho_{mnrs}|m\rangle_{\textrm{cav}}\langle n|_{\textrm{cav}}\otimes|r\rangle_{a_{i}}\langle s|_{a_{i}}\label{denmatrix}
\end{eqnarray}
at time $t_{i}+\tau$, where $|m(n)\rangle_{\textrm{cav}}$ and
$|e(g)\rangle_{a_{i}}$ denote the states of the cavity and the $i$th
injected atom in the cavity QED, respectively, then the reduced
density matrix of the cavity at time $t_{i+1}$ should be
\begin{eqnarray}
\hat{\rho}_{\textrm{cav}}=\texttt{Tr}_{a_{i}}[\hat{\rho}_{\textrm{cav-QED}}]
=\sum_{m,n=0}^{\infty}\sum_{r=\{e,g\}}\rho_{mnrr}|m\rangle_{\textrm{cav}}\langle n|_{\textrm{cav}}\label{denmatrixofcavity}
\end{eqnarray}
taking the trace over the $i$th injected atom. Notice that where we
have neglected the action from the environment of the cavity during
the interval from $t_{i}+\tau$ to $t_{i+1}$. In addition, the
$(i+1)$th atom is prepared in its initial state at time $t_{i+1}$.
Therefore, to simulate the micromaser with the circuit QED system,
it is required that the qubit should be disentanglement from the
resonator at time $t_{i+1}$ to avoid the correlation between the
qubit and the resonator at the beginning of the $(i+1)$th coupling.
By comparing Eq.~(\ref{denmatrix}) with
Eq.~(\ref{denmatrixofcavity}), we can see that the disentanglement
condition is that, during the time interval from $t_{i}+\tau$ to a
time $t_{i}+\tau+\tau_{r}$ before $t_{i+1}$, the qubit should be
relaxed to its ground state as follows:
\begin{eqnarray}
&&|e\rangle_{q}\langle e|_{q}\rightarrow|g\rangle_{q}\langle g|_{q},\nonumber\\
&&|e\rangle_{q}\langle g|_{q}\rightarrow0,\nonumber\\
&&|g\rangle_{q}\langle e|_{q}\rightarrow0,\nonumber\\
&&|g\rangle_{q}\langle g|_{q}\rightarrow|g\rangle_{q}\langle g|_{q}.
\end{eqnarray}
Under this process the density matrix~(\ref{denmatrix}) becomes
\begin{eqnarray}
\hat{\rho}_{\textrm{cir-QED}}=\hat{\rho}_{\textrm{TLR}}\otimes|g\rangle_{q}\langle g|_{q},\label{denmatrix2}
\end{eqnarray}
where we denote
\begin{eqnarray}
\hat{\rho}_{\textrm{TLR}}=\sum_{m,n=0}^{\infty}\sum_{r=\{e,g\}}\rho_{mnrr}|m\rangle_{\textrm{TLR}}\langle n|_{\textrm{TLR}}.
\end{eqnarray}
Clearly, the density matrices $\hat{\rho}_{\textrm{TLR}}$ of the TLR
in the circuit QED and $\hat{\rho}_{\textrm{cav}}$ of the cavity in
the cavity QED have the same form.

Before the beginning time $t_{i+1}$ of the $(i+1)$th coupling, we need to prepare the qubit in its initial state, we denote the state-preparation time is $\tau_{p}$. At time $t_{i+1}$, we switch on the coupling between the qubit and the TLR, repeating the process as described before, we can simulate the micromaser with the circuit QED system.

Now, there are several time scales in the circuit QED system: the interaction time $\tau$, the relaxation time $\tau_{r}$, and the
state-preparation time $\tau_{p}$. From the above discussions we can see that the requirement of these time scales is
\begin{eqnarray}
\tau+\tau_{r}+\tau_{p}\leq t_{i+1}-t_{i}
\end{eqnarray}
for all $i$. In addition, since we neglect the relaxation of the
qubit during the interaction time $\tau$, then the relaxation time
$\tau_{r}\gg\tau$. The preparation of the qubit's state can be
realized by using a classical field. For example, a $2\pi$ pulse can
transfer the qubit from its ground state $|g\rangle$ to excited
state $|e\rangle$. The preparation time $\tau_{p}$ can be much
smaller than other time scales through choosing a sufficiently
strong field. Namely, we approximately have the relation
$\tau+\tau_{r}+\tau_{p}\approx\tau+\tau_{r}$.

In the following, we give a simple estimation of the above time
scales under the current experimental conditions. According to
recent circuit QED experiments~\cite{cQED2}, we take the following
parameters, the resonator frequency $\omega=2\pi\times10$ GHz, the
coupling strength $g=2\pi\times50$ MHz, the cavity decay rate
$\kappa=2\pi\times1$ MHz. The interaction time $\tau\sim10^{-9}$ s
for $g\tau\sim0.05$. The rate to switch on the coupling
$r\approx2\pi\times2$ MHz. Namely, the interaction between the
resonator and the qubit takes place every $t\approx1/r\sim10^{-7}$
s. In other words, the average time
$\overline{t_{i+1}-t_{i}}\approx10^{-7}$ s, where the overline
represents average value. Then the qubit relaxation time $\tau_{r}$
should be of the order of $10$ ns to satisfy the conditions
$\tau+\tau_{r}\leq t_{i+1}-t_{i}$ and $\tau_{r}\gg\tau$. Therefore,
the requirements for the design of the random pulse are $\min
\{t_{i+1}-t_{i}\}\geq \tau+\tau_{r}\sim10^{-8}$ s and
$\overline{t_{i+1}-t_{i}}\approx10^{-7}$ s. We can decrease the
average rate $r$ to enlarge the variable space of the time scale
$\tau_{r}$. As an example, we take the temperatures of the TLS and
the field heat bath as $T=200$ mK ($\beta=2.898$) and $T_{b}=100$ mK
($\beta_{b}=4.797$), respectively. We can calculate the temperature
of the cavity field at thermal equilibrium as
$T_{\textrm{eff}}=100.54$ mK ($\beta_{\textrm{eff}}=4.773$) for the
case $g\tau=0.05$.

\section{\label{Sec:6}Conclusions and discussions}

In conclusion, we have proposed a kind of generalized thermalization
with a single-particle reservoir. This generalized thermalization
uniquely describes the cooling, masering, and thermalization
processes. We have shown our generalized thermalization based on a
micromaser-like system, in which a series of well-prepared TLSs are
injected \textit{randomly} through the cavity. In the absence of the
cavity decay, the cavity can reach an equilibrium with the same
temperature as that of the TLSs. When the cavity is coupled with a
heat bath, at a steady state the cavity can reach a thermal
equilibrium with an intermediate temperature between those of the
heat bath and the TLSs. We have also studied the effect of quantum
coherence on the thermalization. It was found that the quantum
coherence can increase the temperature of the system to be
thermalized at a steady state. We have suggested an experimental
implementation of our generalized thermalization with the
superconducting circuit QED system.

We point out that the present investigations have some potential
values in solid thermodynamical applications. For example, in
circuit QED we can manipulate the steady state of the TLR by
preparing the quantum state of the qubit. When the qubit is prepared
in its ground state, the steady state of the TLR will be in its
ground state. A similar idea has been used to cool a
TLR~\cite{Grajcar2008}. In addition, by preparing the qubit in its
excited state, the TLR can be used to realize a solid laser
on-chip~\cite{Astafiev2007}.

We give some discussions concerning the ergodicity during the
quantum thermalization process. As mentioned in the Introduction,
one of the motivations of our present investigation was to show the
quantum thermalization of a system randomly coupled with a series of
single-particle reservoirs in a time domain. Essentially, the
physical principle at the background is the ergodicity, which
involves the equivalence between the time average and ensemble
average~\cite{Reichl}. For the present case, it has been shown that
the form of the quantum master equation~(\ref{effmasequforkaeqzero})
obtained in the short $\tau$ limit can guarantee the
ergodicity~\cite{Cresser2000,Maassen2003}. Therefore, it is a
correct result that these injected atoms can thermalize the
single-mode cavity field in the micromaser model.

Finally, we emphasize that in this work the injected atoms are
prepared in thermal equilibrium with positive temperature. As is
known, the temperature of two-level atoms can be
negative~\cite{Ramsey}. However, for a harmonic oscillator, it is
impossible to define a negative temperature since the energy of the
harmonic oscillator is finite. Therefore, the discussions in this
paper are restricted within the case that the population $p_{e}$ of
the excited state is smaller than the population $p_{g}$ of the
ground state.

\acknowledgments This work is supported in part by NSFC Grants No.
10935010 and No. 10775048, NFRPC Grants No. 2006CB921205 and No.
2007CB925204.

\end{document}